\documentclass[aps,pre,twocolumn,groupedaddress]{revtex4-1}
\usepackage[version=3]{mhchem}
\usepackage{amsmath,amssymb,graphicx,color}
\usepackage{dcolumn}% Align table columns on decimal point
\usepackage{bm}% bold math
\usepackage[mathlines]{lineno}
\usepackage{epsfig}
\usepackage{graphicx}% Include figure files
\usepackage[cp1251]{inputenc}
\usepackage[english]{babel}

\usepackage{hyperref}
\hypersetup{pdfstartview={FitH},pdfpagemode={UseNone},
            colorlinks,linkcolor=blue, citecolor=blue, urlcolor=blue,
            bookmarks=true, bookmarksopen=true, pdfnewwindow=true}

\newcommand{\rmi}{\mathrm{i}}
\renewcommand{\Re}{\mathop{\mathrm{Re}}\nolimits}
\renewcommand{\Im}{\mathop{\mathrm{Im}}\nolimits}

\begin{document}

\title{Enhanced photonic spin Hall effect with subwavelength topological edge states}

\author{\firstname{A.~P.} \surname{Slobozhanyuk$^{1,2}$},
  \firstname{A.~N.} \surname{Poddubny$^{2,3}$},
  \firstname{I.~S.} \surname{Sinev$^{2}$},
  \firstname{A.~K.} \surname{Samusev$^{2}$},
  \firstname{Y.~F.} \surname{Yu$^{4}$},
  \firstname{A.~I.} \surname{Kuznetsov$^{4}$},
  \firstname{A.~E.} \surname{Miroshnichenko$^{1}$},
  \firstname{Yu.~S.} \surname{Kivshar$^{1,2}$}}

\affiliation{$^1$Nonlinear Physics Center, Australian National University, Canberra ACT 0200,
Australia\\
$^2$Department of Nanophotonics and Metamaterials, ITMO University, St. Petersburg 197101, Russia\\
$^3$Ioffe Physical-Technical Institute of the Russian Academy of Sciences,
St. Petersburg 194021, Russia\\
$^4$Data Storage Institute, A*STAR (Agency for Science, Technology and Research), 138634, Singapore}

%\email[ Corresponding author: ]{a.slobozhanyuk@phoi.ifmo.ru}
%\date{\today}

\begin{abstract}
Photonic structures offer unique opportunities for controlling light-matter interaction, including the photonic spin Hall effect associated with the transverse spin-dependent displacement of light that propagates in specially designed  optical media. However, due
to small spin-orbit coupling, the photonic spin Hall effect is usually weak at the nanoscale. Here we suggest theoretically and demonstrate experimentally, in both optics and microwave experiments, the photonic spin Hall effect enhanced by topologically protected edge states in subwavelength arrays of resonant dielectric particles. Based on direct near-field measurements, we observe the selective excitation of the topological edge states controlled by the handedness of the incident light. Additionally, we reveal the main requirements to the symmetry of photonic structures to achieve {\em a topology-enhanced spin Hall effect}, and also 
analyse the robustness of the photonic edge states against the long-ranged coupling.
\end{abstract}

%\maketitle

\maketitle

Spin-orbit interaction is known to be responsible for the quantum spin Hall effect in topological insulators in solids~\cite{Hasan2010,Bernevig2013}. The study of spin-orbit interaction of light has been a driving force for the development of many
concepts of {\em topological photonics}~\cite{Soljacic2014} dealing with photons instead of electrons and revealing optics 
analogies of the scattering-immune surface states.

In photonics, various designs  mimicking the spin-orbit interaction in solids have been proposed and realized for light~\cite{Wang2009,Fan2012,verbin2013,rechtsman2013,hafezi2013c,Khanikaev2013,ChanNatureComm2014,Slobozhanyuk2015arXiv,Lu2015}.
While the topological photonics focuses on  the {\it quantum} spin Hall effect, the optical analogies of the  {\it classical} spin Hall effect are also well studied~\cite{dyakonov2008spin,RevModPhys_SHE,Bliokh2015}. Namely, a light beam exhibits a transverse shift propagating through an inhomogeneous  medium, and the shift depends on the sign of its circular polarization~\cite{Zeldovich1992}. The concept of photonic spin Hall effect has been predicted for light propagating in free space~\cite{Onoda:PRL:2004,Bliokh_PRL_2006,Bliokh_PRL_2008}, polaritons in microcavities~\cite{Glazov2005,Glazov2007}, and it was demonstrated experimentally in many systems ranging from planar waveguides~\cite{Kwiat2008,Petersen2014} and advanced plasmonic structures~\cite{Hasman2011,ginzburg2013,OConnorNatureComm2014}, to metamaterials \cite{Kapitanova2014} and metasurfaces~\cite{Capasso2013,Yin2013,Kildishev2013,Luo_ADOM_2015,Shalaev_Optica}.

The topological photonics and spin Hall effect for light rely on the same principles of the spin-orbit interaction and nontrivial Berry connection formalism. Nevertheless, with a few exceptions~\cite{Dasha2015}, these studies are focused on quite different phenomena, either the edge states {\it per se} or spin-dependent transverse light propagation.  Here, we bridge these two seemingly distinct fields  of photonics by suggesting the concept of {\em the topology-enhanced spin Hall effect}, namely describe theoretically and demonstrate experimentally the spin Hall effect for light enhanced by the presence of topological edge states.

\begin{figure}[t!]
\protect\includegraphics[width=1.0\columnwidth]{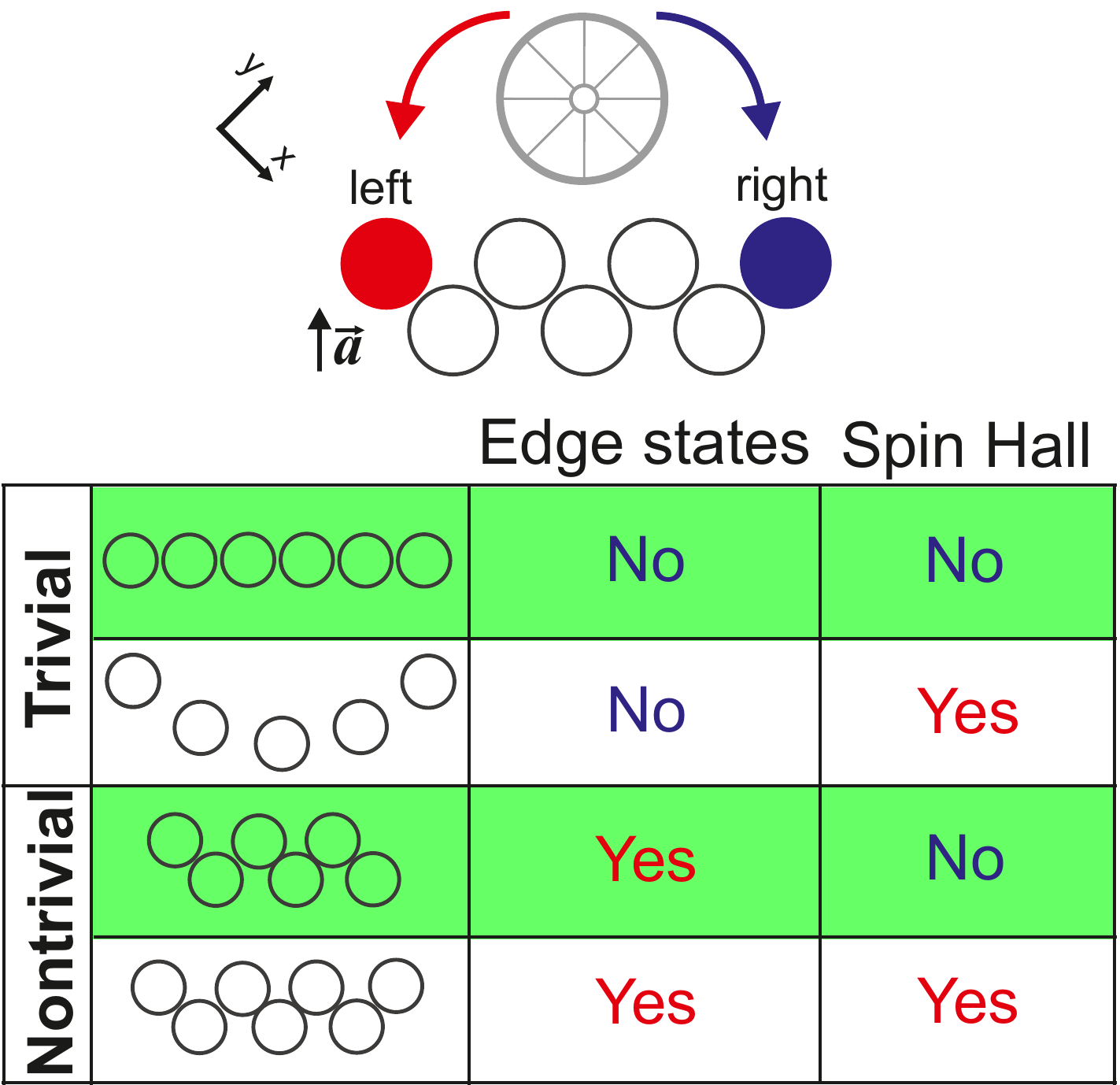}
\caption{\textbf{Concept of the topology-enhanced spin Hall effect in subwavelength particle arrays}. Top: Schematic illustration of the electromagnetic field localization in the left or right part of the nanoparticle array controlled by the polarization handedness. Bottom: Existence of the spin Hall effect and topological edge states for the nanoparticle structures of different geometry. Green shaded area depicts the structures
with the inversion symmetry.
\label{fig:concept}}
\end{figure}

We consider a zigzag array of resonant dielectric particles as the simplest example of a nanostructure supporting nontrivial localized topological states~\cite{Poddubny2014majorana}. Hybridization between  the polarization-degenerate  particle resonances gives rise to a pair of edge states localized at the opposite ends of the array. The topological nature of the states stems from the chiral-symmetric energy spectrum, and it can be characterized by the $\mathbb Z_{2}$ winding number parity~\cite{Slobozhanyuk2015}. In a sharp contrast to our previous studies~\cite{Poddubny2014majorana,Sinev2015,Slobozhanyuk2015,Slobozhanyuk_OPN} restricted  only to the linearly polarized waves, here we present the full polarization spectroscopy that demonstrates non-zero ellipticity of the edge states. This means that the resonant hotspots of the electromagnetic near-field  can be redistributed from the left edge to the right one by changing the handedness of the exciting wave, which is the essence of the photonic spin Hall effect.

\section*{Results}
\label{sec:results}

{\bf Symmetry analysis}. Figure~\ref{fig:concept} illustrates the effect analyzed here for the structures of different geometry and summarizes the concept of this work. Phenomenologically, the zigzag array with an odd number of particles $N$ has $C_{2v}$ symmetry~\cite{Dresselhaus2008}. In the case of the normal incidence of a plane wave with the amplitude $\bm E_{0}$, the transverse shift of the near-field hot spot $\bm \Delta$ can be described by the expression
\begin{equation}
\bm \Delta\propto \bm S\times \bm a,\quad \bm  S={\rm i} \bm E_{0}\times \bm E_{0}^{*}\:
\label{eq:symmetry0}
\end{equation}
where the pseudo-vector $\bm S$ quantifies the circular  polarization of the incident plane wave\cite{Landau04}, and  can interpreted as an electric part of light spin angular momentum density~\cite{Bliokh2011}. 
The vector $\bm a$ is the  $C_{2}$ axis direction (see Fig.~\ref{fig:concept}).  Qualitatively, a lack of the in-plane inversion symmetry  of the zigzag geometry acts as a ratchet, transforming the rotation of the electric field
into the transverse translational shift $\bm \Delta$, and enabling the photonic spin Hall effect. Formally,  the pseudo-vector component $S_z$ transforms in the $C_{2v}$ group according to the same irreducible representation $B_1$ as the vector component $x+y$~\cite{Dresselhaus2008}. This allows the hot spot shift in the direction $\bm e_x+\bm e_y$, linearly proportional to $S_z$. The phenomenological equation Eq.~\eqref{eq:symmetry0} is also analogous to that describing the circular photo-galvanic effect in gyrotropic crystals~\cite{Ivchenko1978,Belinicher1978,IvchenkoPikus}. In the inversion-symmetric structures such as a straight line ($D_{\infty h}$ symmetry ) or a zigzag with an odd number of particles ($C_{2h}$ symmetry), the  effect of type Eq.~\eqref{eq:symmetry0} is impossible since $\bm a=0$, see the green shaded lines of the table in Fig.~\ref{fig:concept}. For the asymmetric $C_{2v}$ structures such as an arc or a zigzag with an odd number of particles  where $\bm a \ne 0$, the effect is allowed by symmetry (white/unshaded lines of the table). However, contrary to the arc, the zigzag with an odd number of particles also supports nontrivial edge states. Our main result is the prediction of the substantial enhancement of the photonic spin Hall effect by employing the intrinsic features of these spin-dependent topologically nontrivial edge states.

{\bf Coupled-mode theory}. To illustrate the emergence of the photonic spin Hall effect at the topologically nontrivial
edge states, we start with the general Hamiltonian for a zigzag array written in the form,
\begin{multline}
H=\sum\limits_{j,\nu}\hbar \omega_{0}a_{j\nu}^{\dag}a^{\vphantom{\dag}}_{j\nu}+\\
\sum\limits_{\langle j,j'\rangle,\nu,\nu'}a_{j\nu}^{\dag}V^{(j,j')}_{\nu\nu'}a^{\vphantom{\dag}}_{j'\nu'}+
\sum\limits_{\langle\langle j,j'\rangle\rangle,\nu,\nu'}a_{j\nu}^{\dag}W^{(j,j')}_{\nu\nu'}a^{\vphantom{\dag}}_{j'\nu'}
\:, \label{eq:H}
\end{multline}
where $\omega_{0}$ is the resonance frequency, the indices $j$ and $j'$ mark the particles, and
$\langle j,j'\rangle$ (\,$\langle\langle j,j'\rangle\rangle$\,) are the first (second) nearest neighbors in the array.
The symbol $a_{j\nu}$ stands for the annihilation operator for the states with the polarization $\nu$ at the $j$-th particle. We restrict our analysis by the states excited at the normal light incidence upon the zigzag plane $(x,y)$. This includes
$x$-polarized and $y$-polarized  electric/magnetic dipole modes or $xz$- and $yz$-polarized quadrupole modes. Hybridization between each given kind of modes at different particles  can be described by the same Hamiltonian~\eqref{eq:H}~\cite{Slobozhanyuk2015}.
The coupling matrices $V_{\nu\nu'}$ have the form
\begin{equation}
V^{(j,j')}_{\nu\nu'}=t^{(1)}_{\parallel} \bm  e^{(j,j')}_{\parallel}\otimes\bm e^{(j,j')}_{\parallel}+
t^{(1)}_{\perp} \bm  e^{(j,j')}_{\perp}\otimes\bm e^{(j,j')}_{\perp}\:,
\end{equation}
where $ \bm  e^{(j,j')}_{\parallel}$ ($\bm  e^{(j,j')}_{\perp}$) is the in-plane unit vector parallel (perpendicular) to the link vector  $\bm r_{j}-\bm r_{j'}$ and $t^{(1)}_{\parallel}$ and $t^{(1)}_{\perp}$ are the coupling constants  for the  modes, co-polarized and cross-polarized with respect to $\bm r_{j}-\bm r_{j'}$.  In the case of the near-field quadrupole-quadrupole interaction $t_{\parallel}/t_{\perp}=-4$, and for the dipole-dipole interaction, $t_{\parallel}/t_{\perp}=-2$.  Within the nearest-neighbor approximation ($W\equiv 0$), the edge states in the right angle zigzag  have strictly $x$ or $y$ polarization. As such, they are excited  with the same amplitude by the right- and left-circularly polarized waves, and the photonic spin Hall effect is not revealed.

\begin{figure}[b!]
\protect\includegraphics[width=1.0\columnwidth]{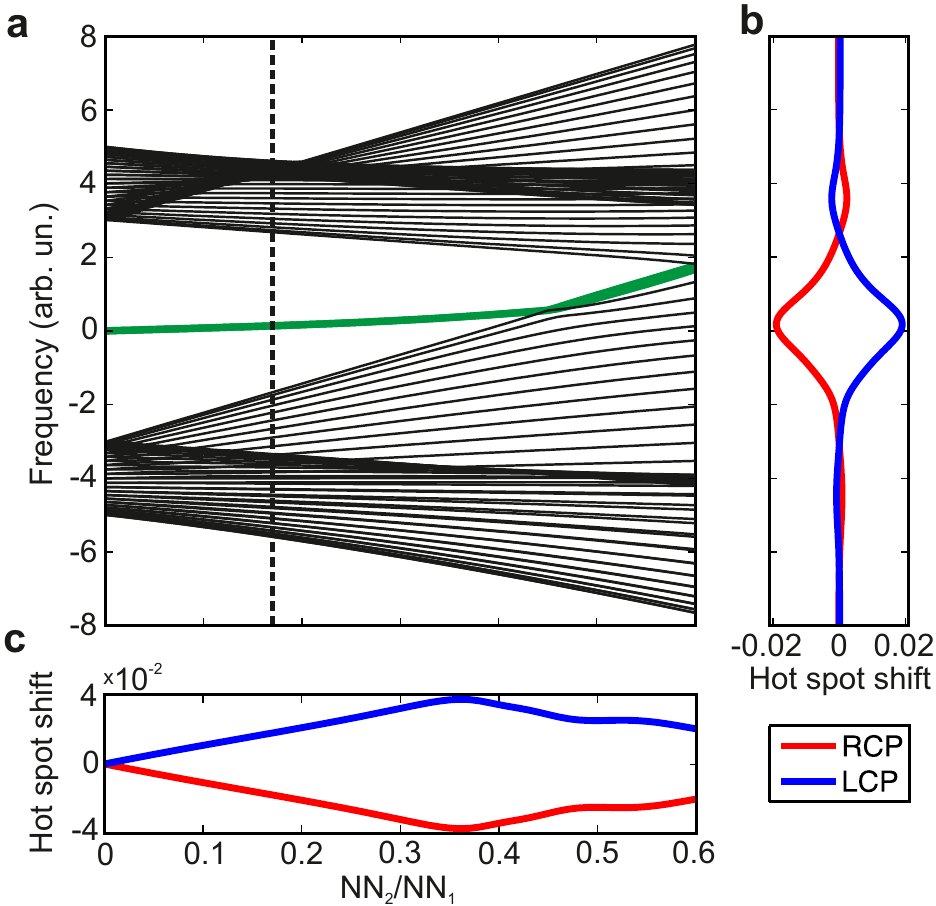}
\caption{ \textbf{Microscopic origin of the spin Hall effect. a} Energy spectrum as a function of the relative second-nearest neighbor coupling strength. Energies of two degenerate edge states are shown in green. {\bf{b}} Near field hot spot shift from the zigzag center as a function of the exciting wave frequency at $NN_{2}/NN_{1}=0.18$ (dashed vertical line in panel {\bf{a}}).
{\bf{c}} Near field hot spot shift at $\omega=0$ as function of the  relative second-nearest neighbor coupling strength. %(d) Excited near field spatial distributions $|p_{j}|^{2}$ for $\omega=0$ and $NN_{2}/NN_{1}=1/\sqrt{2}^{5}$.
The calculation parameters are given in text.
\label{fig:1b}}
\end{figure}

\begin{figure*}[t]
\includegraphics[width=2.0\columnwidth]{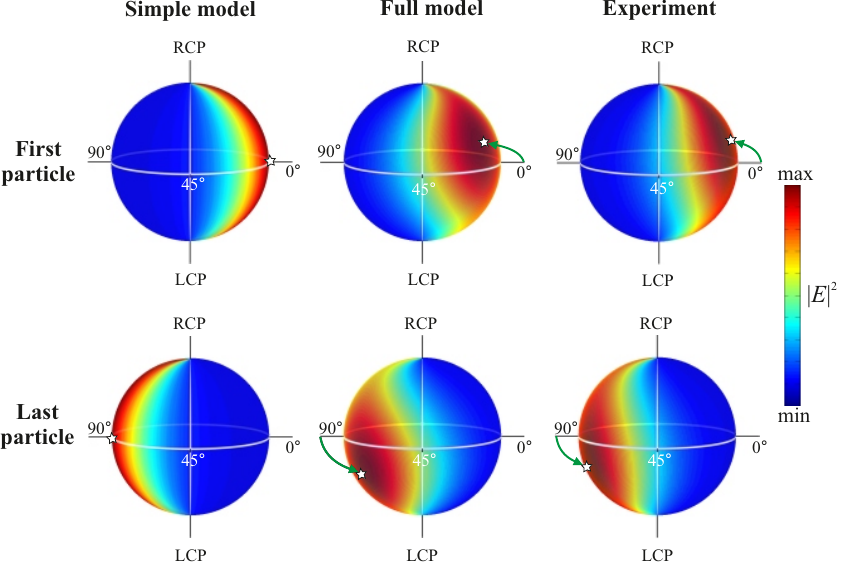}
%\protect\includegraphics[width=2.0\columnwidth]{Fig_3a_Final_1609}
\caption{\label{Fig1}\textbf{Excitation-dependent polarization spectroscopy}.  Electric field intensity near the first particle (top line) and last particle (bottom line) of the zigzag array for different polarizations of incident wave shown on the Poincare sphere. Three panels correspond to nearest-neighbor model, full numerical and experimental results. White stars indicate the maxima of the field.}
\end{figure*}

The coupling terms describing the interaction of the second neighbors (${NN_{2}}$) make a crucial difference between the considered model and that discussed earlier~\cite{Slobozhanyuk2015}; they mix both $x$- and $y$- polarized modes thus providing the microscopic explanation for the physical origin of the photonic spin Hall effect in this system. The matrix describing the second-neighbor couplings  $W^{(j,j')}_{\nu\nu'}$ has the same structure as $V^{(j,j')}_{\nu\nu'}$, and it is characterized by the constants  $t^{(2)}_{\parallel}$ and $t^{(2)}_{\perp}$.    This can be illustrated by introducing the circularly polarized basis $\bm e_{\pm}=(\bm e_{x}+\rmi \bm e_{y})/\sqrt{2}$. In this basis, the matrix elements of the second-order interaction read
 \begin{equation}
 \langle \pm |W|x\rangle=\mp\rmi  \langle \mp |W|y\rangle=\frac{\sqrt{2}}{4}[t^{(2)}_{\parallel}+t^{(2)}_{\perp}\pm \rmi (t^{(2)}_{\parallel}-t^{(2)}_{\perp})]\:.
 \label{eq:SO}
 \end{equation}
Equation~\eqref{eq:SO} shows that for complex couplings $t^{(2)}_{\parallel}$ and $t^{(2)}_{\perp}$ the matrix elements of the Hamiltonian between given linearly-polarized state and two right and left circularly polarized (RCP and LCP) states have different absolute values. This means that the two edge states acquire non-zero and opposite ellipticity. The linear-to-circular couplings~\eqref{eq:SO} can be understood as an effective spin-orbit interaction in this system~\cite{Amo2015}.

 Figure~\ref{fig:1b} shows the effect of the {$NN_{2}$} couplings on the energy spectrum. In Fig.~\ref{fig:1b}(a), we demonstrate the role of the {\it retarded} coupling on the spectrum. To do this, we diagonalize the model Hamiltonian~\eqref{eq:H} for $N=51$ particles with $t^{(1)}_{\parallel}=4$, $t^{(1)}_{\perp}=-1$, $t^{(2)}=t^{(1)}(1+0.2 {\rm i} )NN_{2}/NN_{1}$ and $\omega_{0}=0$. The imaginary term in the couplings, rendering the Hamiltonian non-Hermitian, has been added to reflect the retarded character of the interaction. Shown in
Fig.~\ref{fig:1b}(a) are the real parts of the modes eigenfrequencies as functions of the relative strength of the next-to-nearest-neighbor couplings $NN_{2}/NN_{1}$. This calculation demonstrates that the edge states are robust against the long-ranged couplings. Namely, for $t^{(2)}=0$ the spectrum is symmetric with respect to the central frequency and possesses a pair of polarization degenerate eigenstates localized at the opposite edges of the structure. For non-zero $t^{(2)}$ the Hamiltonian loses the chiral symmetry to the sublattice inversion and the spectrum becomes asymmetric. However, the edge states persist and remain degenerate. More importantly, the states become elliptically polarized for $t^{(2)}\ne 0$.  To demonstrate the resulting spin Hall effect, we calculate the response of the system to the left- or right- circularly polarized waves, determined by the equation
\begin{equation}
(H-\omega-\rmi \gamma)p= E_0,
\end{equation}
where $p$  is the  vector of the dipole (quadrupole) moments of the $2N$ length, and $E_0$ stands for the exciting plane wave;
$E_{0,x}/E_{0,y}=\pm i$, $E_{0,x}=1/\sqrt{2}$ for RCP (LCP) excitation at each particle; $\gamma$ is the resonance damping. The hotspot shift from the center of the zigzag can be calculated as
\begin{equation}
\delta=\frac{\sum\limits_{j=1}^{N}\sum\limits_{\nu}|p_{j,\nu}|^{2}s_{j}}{\sum\limits_{j=1}^{N}\sum\limits_{\nu}|p_{j}|^{2}},\quad s_{j}=\frac{j-N-1/2}{N/2}\:.
\end{equation}
The calculated spectra $\delta(\omega) $ are shown in Fig.~\ref{fig:1b}(b) for the particular case of $NN_{2}=NN_{1}=1/(4\sqrt{2})\approx 0.18$, corresponding to the quadrupole-quadrupole coupling, and $\gamma =1$.  The hotspot shifts to the left or right edge depending on the circular polarization sign and the shift has a resonance at the edge states  frequency. There also exists a bulk contribution to the shift when the excitation frequency is within the band of propagating states rather than is tuned to the edge states. Our analysis shows that  the bulk term is weaker than the edge term when the couplings are significantly retarded, $\Im t^{(2)}\sim \Re t^{(2)}$, which is the case for the actual experimental structures. Panel (c) demonstrates how the hotspot shift increases with the second-neighbor coupling strength when excited at the resonance ($\omega=0$).

Next, we confirm our analytical predictions of the photonic spin Hall effect in the zigzag array of dielectric particles
by rigorous full-wave numerical simulations and the experiments conducted for both microwave and optical frequency range.

\begin{figure}[t!]
\protect\includegraphics[width=1.0\columnwidth]{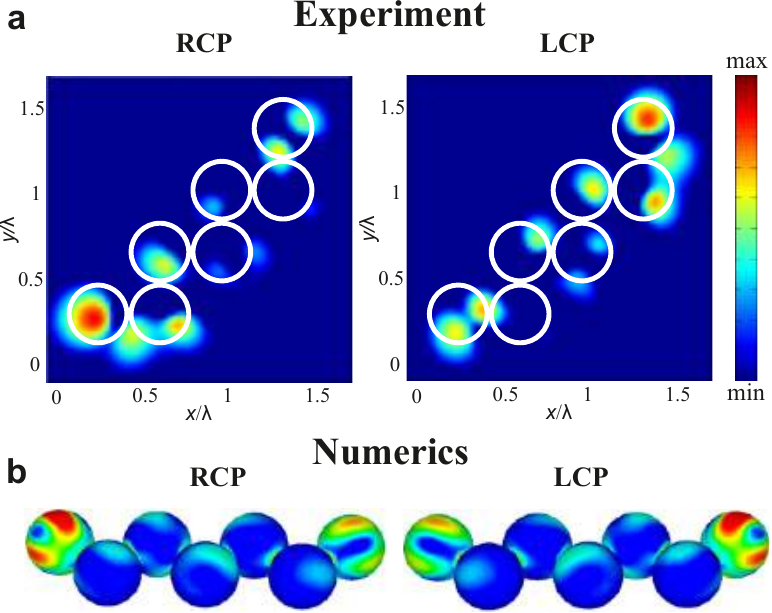}
\caption{\label{Fig3b}\textbf{Observation of the topology-enhanced photonic spin Hall effect}. \textbf{a} Experimentally measured near-field images for the right- and left- circular polarizations.\textbf{b} Numerically calculated amplitudes of the electric field at the surface of the spheres for both right- and left- circularly-polarized incident waves at the frequency of the quadrupole resonance of a single dielectric particle.}
\end{figure}

{\bf Numerical simulations.} We study a closely packed zigzag array of dielectric nanospheres with the permittivity $\varepsilon = 15$ (Fig.~\ref{fig:concept}), and focus on the edge states at the magnetic quadrupole resonance~\cite{Slobozhanyuk2015}. The polarization-dependent  excitation of the edge states can be shown conveniently on the Poincare sphere. In particular, Fig.~\ref{Fig1} presents the electric field intensity at the first (upper row) and last (lower row) particles as a function of the polarization of the incident wave. In the nearest-neighbor approximation, the field intensity remains the same for the right- and left- circularly polarized excitations; i.e. the map has a mirror symmetry with respect to the equator of the Poincare sphere. In order to include all possible couplings in our analysis, we perform the full-wave numerical simulations of the electromagnetic field by using the finite element method (see Methods for details). A change of the calculated maps along the equator reflects the sensitivity of the edge mode to the linear polarization. Clearly, the first particle is preferentially excited by the $x$-polarized light, and the last particle --- by the $y$-polarized light. A variation along the meridian encodes the effects of the circular polarization, revealing the photonic spin Hall effect. The calculation demonstrates that the maximum, shown by a star, is shifted to the north pole of the sphere in the case of first particle (top central panel of Fig.~\ref{Fig1}), or the south pole of the Poincare sphere in the case of last particle (bottom central panel of Fig.~\ref{Fig1}). This means that  the near field is preferentially localized at the  first particle for the right circularly polarized excitation, and on the last particle, for the left circular polarization.

{\bf Experimental verification at microwaves.} For the proof-of-concept experimental demonstration, first we study the effect in the microwave range for an array of MgO-TiO$_{2}$ ceramic spheres characterized by the dielectric constant of 15 and small dielectric loss factor measured in the 4-10~GHz frequency range. For more details about the sample and experimental setup see Methods. Figure~\ref{Fig3b}a shows the experimentally measured near-field images in the zigzag structure for the right- and left- circularly polarized incident waves at the frequency of the magnetic quadrupole resonance of a single dielectric sphere. We observe that the electric field has the maxima either at the left or right edges of the array which is controlled by the handedness of the incident wave. This is a direct observation of  the photonic spin Hall effect for the subwavelength edge states in such an array. Furthermore, we calculate the amplitude of the electric field induced in the zigzag array for the right- and left- polarized incident waves (see Fig~\ref{Fig3b}b). We observe a good agreement between the experimental and numerical results (cf. Fig~\ref{Fig3b}a and Fig~\ref{Fig3b}b).

\begin{figure*}[t!]
\includegraphics[width=2.0\columnwidth]{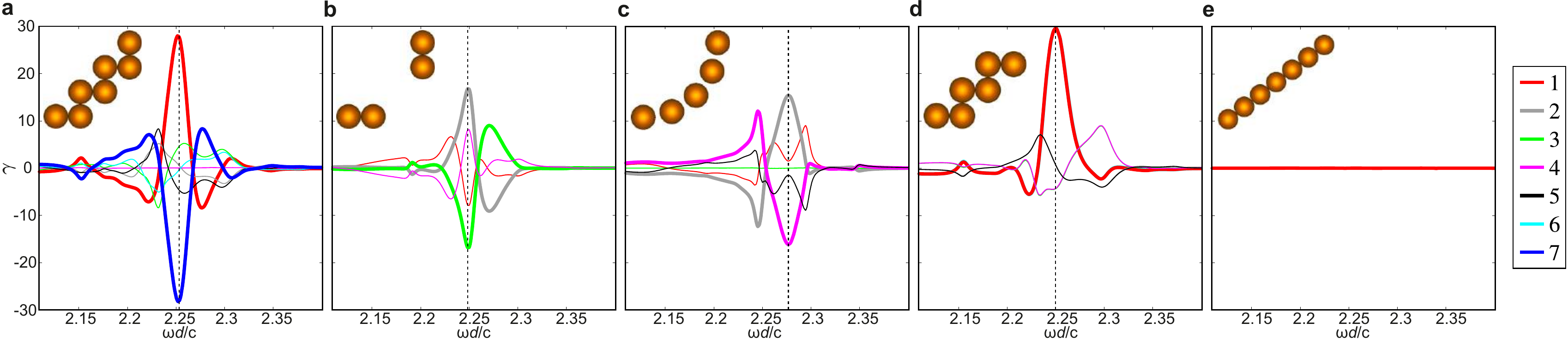}
%%\protect\includegraphics[width=2.05\columnwidth]{Figure2_NEW}
\caption{\textbf{Spin Hall effect in the structures with different symmetries}. Spectral dependence of the $\gamma$ near the centre of each particle of the structures with different geometry (shown as an insets) calculated near the quadrupole resonance frequency of the single sphere. Maximum of the $\gamma$ for each structure marked as a black dashed line.
 }\label{Fig2}
\end{figure*}

We also perform the full excitation-dependent polarization spectroscopy. The measured  intensity of the electric field above  the first particle (top line) and the last particle (bottom line) as a function of incident wave polarization is shown in the last panel of Fig~\ref{Fig1} and confirms the full wave numerical simulations in the central panel of Fig~\ref{Fig1}.

{\bf Near-field analysis.} Further information on the spin Hall effect is gained from the near-field maps in the structures with different symmetry. The calculated frequency-dependent results  are shown in Fig.~\ref{Fig2}. The strength of the effect can be quantified by  the difference $\gamma=I_{RCP}-I_{LCP}$, where $I_{RCP}$ and $I_{LCP}$ are the electric field intensities for two corresponding circular polarizations. Figure~\ref{Fig2} shows the spectral dependence of the $\gamma$ near the centre of each particle for the different structures  consisting of identical resonant dielectric particles in the spectral vicinity of the magnetic quadrupole resonance.
All three structures in  Figs.~\ref{Fig2}a,b,c, the zigzag array, two orthogonal  dimers, and the arc,  have the same $C_{2v}$ symmetry that allows for the spin Hall effect. However, the value of $\gamma$ for the zigzag array (Fig~\ref{Fig2}a) is approximately twice higher than for two  dimers (Fig~\ref{Fig2}b) or for the arc (Fig~\ref{Fig2}c). This illustrates the enhancement of the  effect at the elliptically polarized topological edge states. For comparison we also show the response for the zigzag array  with even  number of particles (Fig~\ref{Fig2}d) and for the straight line (Fig~\ref{Fig2}e). These structures do not exhibit the spin Hall response due to the presence of the center of inversion symmetry.

\begin{figure*}[t!]
\includegraphics[width=2\columnwidth]{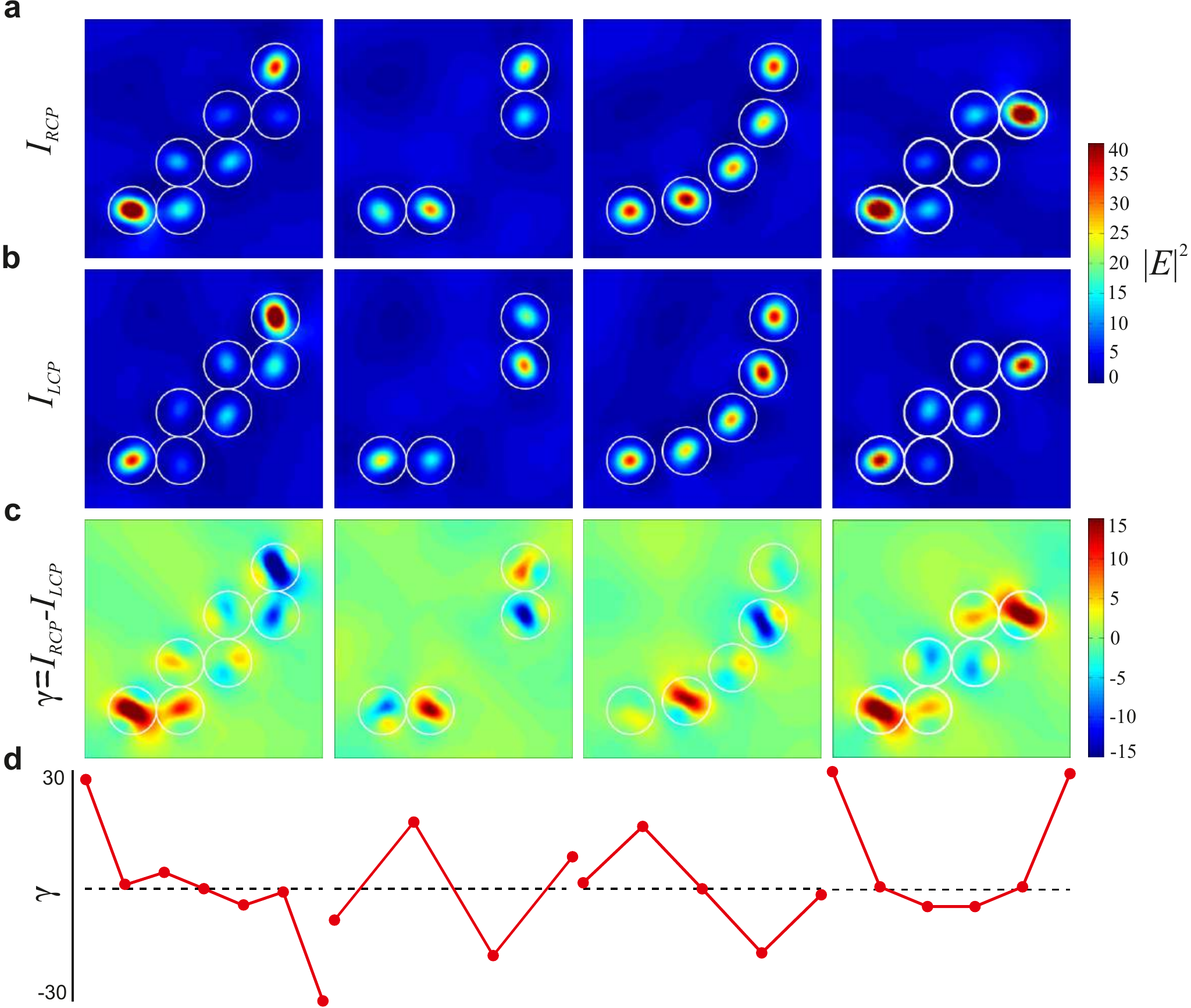}
\caption{\textbf{Near-field mapping of the resonant modes.} \textbf{a,b} Intensity of the electric field induced in the discrete arrays for the right- and left- circularly polarized excitation. \textbf{c} Spatial dependencies of the strength of the spin Hall effect $\gamma$. \textbf{d} Dependence of the  strength of the spin Hall effect $\gamma$ measured above particle's centre as a function of the particle number. The images are calculated at  the frequencies corresponding to the maximum of the spin Hall effect  (see the spectra in Fig.~\ref{Fig2}).}
\label{Fig3}
\end{figure*}

Actual near-field maps for these structures are shown in Fig.~\ref{Fig3}. Figure~\ref{Fig3}a and Fig.~\ref{Fig3}b present the  electric field intensity at the frequencies where the value of $\gamma$ is at maximum, for two different handednesses of the excitation. The spatial dependence of the $\gamma$, obtained as a difference of the images in Fig.~\ref{Fig3}a and Fig.~\ref{Fig3}b,  is shown in the panels (c) and (d). In agreement with the spectral dependence in Fig.~\ref{Fig2}  the spin Hall effect  is significantly more pronounced for the zigzag array (Fig.~\ref{Fig3}, first column) in comparison with the orthogonal dimer structure (second column) and the arc structure (third column). For the zigzag array with  even number of particles the spin Hall effect is absent, i.e. the distribution of the near field has mirror symmetry with respect to the zigzag center for both circular polarizations. However, the amplitude of the field enhancement at the edges with respect to the center is different for right- and left- circular polarizations. Therefore, the parameter $\gamma$ in Fig.~\ref{Fig2}d and in the last column of Fig.~\ref{Fig3} is not zero; it reflects the stronger  coupling of the RCP excitation to the edge states.

\begin{figure*}[t]
\includegraphics[width=1.8\columnwidth]{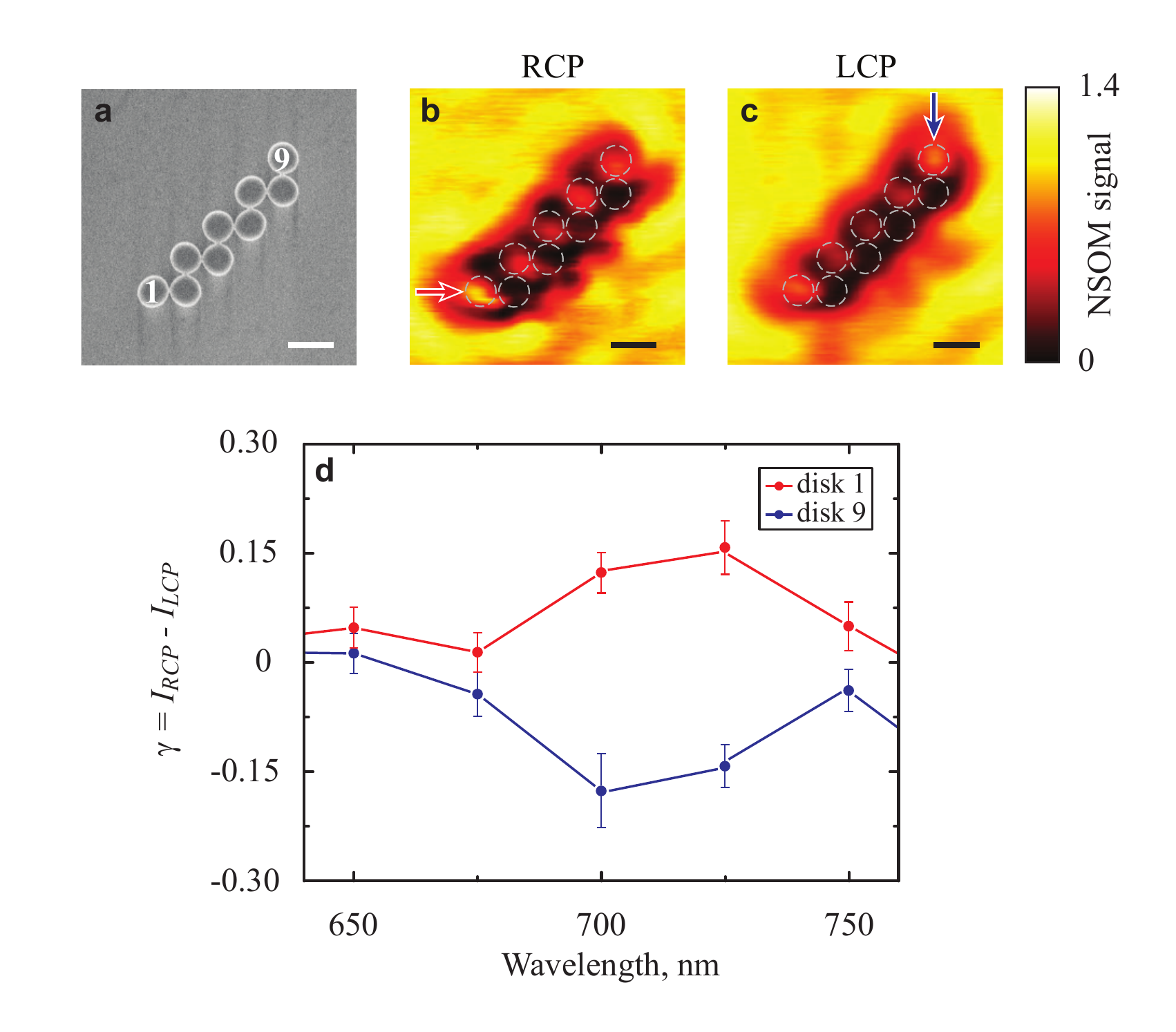}
\caption{{\textbf{Experimental observation of the topology-enhanced photonic spin Hall effect in optics.} \textbf{a} Scanning electron microscopy image of the zigzag array of Si nanodisks. \textbf{b,c} Near-field optical images measured for (b) right and (c) left circularly polarized excitation demonstrating the photonic spin Hall effect. The scale bars in (a-c) represent 500~nm. (d) Experimentally obtained spectral dependencies of the parameter $\gamma$ for the edge nanodisks 1 and 9 marked with the numbers on the SEM image in the panel (a) and with arrows in the panels (b,c).}
\label{Fig7}}
\end{figure*}

{\bf Optics experiments.} In order to demonstrate the predicted features of the photonic spin Hall effect in the optical range, we fabricate a zigzag array composed of nine silicon (Si) nanodisks placed on a glass substrate (see Fig.~\ref{Fig7}a). As was shown previously, Si nanoparticles support strong (dipole, quadrupole, and higher-order) electric and magnetic resonances which spectral positions are tunable with the variation of the particle size~\cite{Evlyukhin2010,Garcia2011,Kuznetsov2012,Bakker2015}. Here, the fabricated nanodisk diameter is around 350~nm, with the neighboring nanodisks touching each other due to slight shape imperfections (Fig.~\ref{Fig7}a).

The effect of the field localization at the edges of the array controlled by the handedness of the excitation light is confirmed experimentally by using aperture-type near-field scanning optical microscope (see Methods for more details about the sample fabrication and experimental setup). The near-field maps measured for the right- and left-circularly polarized light at the wavelength 725~nm are shown in Figs.~\ref{Fig7}b,c. These maps demonstrate the localization of the measured near-field signal at both left or right edges of the array depending on the polarization handedness of the probing beam. In agreement with the results obtained in the microwave range (Fig.~\ref{Fig3b}), we observe the enhanced near-field signal at the left nanodisk for the right circularly polarized excitation and at the right nanodisk for the left circularly polarized excitation (see Figs.~\ref{Fig7}b,c, respectively). The experimental spectral dependence of the parameter $\gamma$ is presented in Fig.~\ref{Fig7}d. $I_{RCP}$ and $I_{LCP}$ are obtained for each wavelength by averaging the near-field signal over the area corresponding to the edge nanodisks of the zigzag array (marked with dashed circles in Figs.~\ref{Fig7}b,c) for the corresponding polarization. The absolute value of the $\gamma$ parameter demonstrates resonant behaviour and reaches the maximum between 680 nm and 730 nm. This matches the spectral region of the existence of the topological edge state for the linear polarization corresponding to the magnetic dipole resonance of a single nanodisk. The experimentally observed switching of the field localization between the edges of the zigzag array for different polarization handedness of the excitation wave provides full evidence of the photonic spin Hall effect in the topologically nontrivial structures in the visible spectral range.

\section*{Discussion}
\label{sec:Discussion}

Our results reveal the specific features of the photonic spin Hall effect mediated by the excitation of the subwavelength topological edge states. We consider one-dimensional arrays of resonant dielectric particles, where both the photonic spin Hall effect and edge states can be independently switched on and off  by judicious engineering  of the array shape. We indicate peculiarities of the structure symmetry required for enhancement of the photonic spin Hall effect via topological properties. Namely,  we demonstrate experimentally and theoretically that the topological edge states in the zigzag array with odd number of particles  substantially enhance the photonic spin Hall effect. By employing near-field scanning techniques, we directly map selective excitation of either right or left edge of the zigzag array controlled by the handedness of an incident plane wave, for both microwave and optical experimental structures. The demonstrated phenomena can be useful for manipulation and routing of light at subwavelength scale as well as for mimicking spin currents from solid state physics in optics.

\section*{Methods}\label{sec:methods}
\textbf{Numerical modelling}. Numerical simulations are performed by using the frequency domain solver of the commercially available software CST Microwave Studio. We model the finite structures in air. The PML boundary conditions are used. For each structure, an iteration of adaptive mesh refinement steps is performed. The modelled structures consist of dielectric spheres with the following parameters: the material of the spheres $\varepsilon = 15$, the diameter of the spheres was equal to $\lambda/2.8$, where the  $\lambda$ is the incident wavelength, the period of the zigzag structures (Fig.~\ref{Fig2}a,d), dimer structure (Fig.~\ref{Fig2}b) and straight line (Fig.~\ref{Fig2}e) is equal to the sphere diameter. In the case of the arc structure (Fig.~\ref{Fig2}c) the period is optimized in order to achieve the same spatial dimension as the zigzag structure (see Fig.~\ref{Fig2}a).

The arrays are assumed to be excited by the circularly polarized plane wave, and the electric field amplitude on the surface (Fig.~\ref{Fig3b}b) or electric field intensity (Fig.~\ref{Fig1}, Fig.~\ref{Fig2} and Fig.~\ref{Fig3}) right behind the spheres is analyzed. In order to plot the electric field intensity in the vicinity of the first particle for different polarizations on the Poincare sphere (Fig.~\ref{Fig1}, top middle sphere), we calculate the complex electric field excited at the first particle for the linearly polarized plane waves with $\theta = 0^0$ and $\theta = 90^0$. After that, the electric field intensity for all possible cases of incident polarization is calculated by using the linear combination of these two orthogonal excitations.

\textbf{Microwave experiments}. For the experimental proof-of-principle verification at microwaves, we use ceramic spheres with a dielectric constant of 15. The sphere radius is 7.5~mm and the period is 15~mm. To fasten the particles together we use a special holder made of a styrofoam material with a dielectric permittivity of 1. To approximate the plane wave excitation, a rectangular horn antenna is used. It is connected to the transmitting port of a vector network analyzer (Agilent E8362C). The sphere is located in the far-field of the antenna (at approximately 2.5 m). The near field measurements are performed for the frequencies within 6-8~GHz frequency band. We use an automatic mechanical near field scanning device and an electric field probe connected to the receiving port of analyzer. The probe is oriented normally with respect to the interface of the structure. The near field is scanned at a 1 mm distance from the back interface of the zigzag array to avoid the probe and the sample touching. In order to determine the structure response to an  RCP or an LCP excitation, we collect the amplitude and phase of the field excited near the structures by a plane wave with the linear polarization $\theta = 0^0$ and $\theta = 90^0$. The resulting experimental maps in Fig.~\ref{Fig3b}a are obtained by combining the measured complex fields for two orthogonal excitation polarizations in MATLAB.

\textbf{Optics experiments}.  To confirm our predictions in the optical range, we fabricate zigzag arrays of  Si nanodisks using a standard top-down nanofabrication procedure.  A 165~nm thick amorphous silicon (Si) film is deposited on a glass substrate using inductively couple plasma chemical vapor deposition (Plasmalab System 380, Oxford Instruments). Electron beam lithography (Elionix, 100kV) is carried out by using Hydrogen silesquioxane (HSQ, Dow Corning, XR-1541-006) as a resist.  Unexposed HSQ is removed with Tetramethylammonium hydroxide (TMAH, 25\%).  The sample is then etched using an inductively coupled plasma etcher (Plasmalab System 100, Oxford) to create Si zigzag arrays of nanodisks on the glass substrate. The fabricated structures are composed of 165~nm tall Si nanodisks with the diameter of around 350~nm with the neighboring disks of the chain touching each other. The samples are visualized by means of Scanning Electron Microscope (SEM, DA300, FEI).

The near-field profiles are measured with a near-field scanning optical microscope (NSOM, AIST-NT TrIOS setup, AIST-NT, Inc). In the experiments, the sample is illuminated from the side of the substrate by either circularly or linearly polarized light, focused to a spot size of approximately 10 microns on the sample surface with a lens (focal length 3~cm). We use a supercontinuum source (Fianium WhiteLase SC400-6) combined with a tunable band-pass filter (Fianium SuperChrome) yielding a beam with a spectral width of 10 nm and tunable central wavelength to excite the sample. The NSOM mapping is carried out in the constant-height mode at the elevation of 100~nm above the surface of the nanodisks to avoid the topography-induced artifacts in the measured signal. The signal is collected by an Al-coated polymer fiber probe with a focused ion beam milled aperture of 75~nm (Lovalite P200).

\section*{Acknowledgements}\label{sec:Acknowledgements}

The authors are grateful to K. Bliokh for many highlighting comments and suggestions, D. Powell for useful discussions, V. Valuckas (DSI) for SEM imaging and N. Emani (DSI) for a help with EBL layout and silicon film deposition. This work was supported by the Australian Research Council, the Government of the Russian Federation (Grant 074-U01), and the Russian Foundation for Basic Research (grant 15-32-20866). ANP acknowledges a support of the Dynasty Foundation. YFY and AIK were supported by the DSI core funds 	
and A*STAR SERC Pharos program (grant No. 152 73 00025).

\section*{Author Contributions}\label{sec:Author Contributions}
All authors contributed extensively to the work presented in this paper.

\bibliography{ZigZag_plus_SpinHall}

\end{document}